# Memristive phase switching in two-dimensional crystals

**Masaro Yoshida[1]\*, Ryuji Suzuki[1], Yijin Zhang[1], Masaki Nakano[1], Yoshihiro Iwasa[1,2]**

[1]*Quantum-Phase Electronics Center (QPEC) and Department of Applied Physics,*

*The University of Tokyo, Tokyo 113-8656, Japan*

[2]*RIKEN Center for Emergent Matter Science (CEMS), Wako 351-0198, Japan*

\*Corresponding author: masaro-yoshida@mp.t.u-tokyo.ac.jp

**Scaling down materials to an atomic-layer level produces rich physical and chemical properties as exemplified in various two-dimensional (2D) crystals extending from graphene, transition metal dichalcogenides to black phosphorous. This is caused by the dramatic modification of electronic band structures. In such reduced dimensions, the electron correlation effects are also expected to be significantly changed from bulk systems. However, there are few attempts to realize novel phenomena in correlated 2D crystals. Here, we report memristive phase switching in nano-thick crystals of 1T-type tantalum disulfide (1T-TaS$_2$), a first-order phase transition system. The ordering kinetics of the phase transition was revealed to become extremely slow as the thickness is reduced, resulting in an emergence of metastable states. Furthermore, we realized the unprecedented memristive switching to multi-step non-volatile states by applying in-plane electric field. The reduction of thickness is essential to achieve such non-volatile electrical switching behavior. The thinning-induced slow kinetics possibly makes the various metastable states robust and consequently realizes the non-volatile memory operation. The present result indicates that 2D crystal with correlated electrons is a novel nano-system to explore and functionalize multiple metastable states which are inaccessible in its bulk form.**

## Introduction

The memristors are switchable resistors with multiple non-volatile memory function (*1,2*). Such devices are predicted to play key roles in developing neuromorphic circuits, ultradense information storage and other applications. One of the promising routes to realize memristive responses is to utilize materials with first-order phase transitions, sometimes called as phase-change materials (*3,4*), providing us with an opportunity to electrically control resistive states. As for the candidate materials for memristors, major targets have been oxides for around 50 years (*5,6*). However, the potential of the larger class of non-oxides should be also examined. Since all the memristors have been realized in nanoscale devices (*7-9*), two-dimensional (2D) crystal (*10-16*) with first-order phase transitions are highly promising. We chose tantalum disulfide with 1T polytype (1T-TaS$_2$), which is a well-known layered system with first-order charge-density-wave (CDW) phase transitions (*17-21*).

1T-TaS$_2$ (Fig. 1A), a correlated transition metal dichalcogenide (TMD), undergoes successive first-order phase transitions on cooling; one is from an incommensurate (IC) to a nearly-commensurate (NC) CDW phase at 350 K, and the other is from the NC to a commensurate (C) CDW phase at 180 K (*17*). In the CCDW phase, thirteen Ta atoms form a David-star cluster as shown in Fig. 1A, where the Mott state is simultaneously developed (*18*). The neighboring NCCDW phase is a hexagonal array of CCDW domains (*19*). Only upon warming between 220 K and 282 K, there appears another NCCDW phase, consisting of striped triclinic CCDW domains (*19*). In both NCCDW phases, the electronic conduction derives from the mobile carriers in domain boundaries, or the discommensuration regions. The schematic pictures of the three CCDW, NCCDW and ICCDW phases are illustrated in Fig. 1B.

The uniqueness of the nano-thick crystal of 1T-TaS$_2$ is its unexpected thickness dependence of the CDW phase transition (*22*). As shown in Fig. 1C, the NCCDW-CCDW transition, which is clearly observed in thicker crystals, is completely absent in the 24-nm-thick crystal while the ICCDW-NCCDW transition remains down to the 7-nm-thick crystal. Figure 1D is the summarized electronic phase diagram on cooling at 1 K/min as a function of thickness, showing an abrupt suppression of the NCCDW-CCDW transition and the stabilization of the super-cooled NCCDW state in crystals thinner than 40 nm. Another group also reported that the NCCDW-CCDW transition vanishes suddenly with reducing thickness (*23*). In terms of statistical mechanics, a continuous change of a physical parameter like thickness cannot result in such a sudden disappearance of long-range ordering. We, therefore, need to employ another view point; kinetics (*24*), to understand and functionalize this peculiar first-order phase transition in the nano-system.

## Results

**Thickness dependent ordering kinetics in 1T-TaS$_2$**

In order to clarify the kinetics of the first-order phase transition in 1T-TaS$_2$ nano-thick crystals, we investigated the cooling rate dependence in many crystals with different thicknesses. Figure 2A displays the temperature ($T$) dependence of resistivity ($\rho$) with different cooling/warming rate for the 51-nm-thick crystal. Upon cooling at 5 K/min, the NCCDW-CCDW phase transition was completely suppressed and the super-cooled NCCDW state was stabilized down to the lowest temperature. In marked contrast, the 15-nm-thick crystal did not exhibit the NCCDW-CCDW phase transition and the super-cooled NCCDW state was stabilized even with the slow cooling rate of 0.3 K/min as shown in Fig. 2B, which reflects its slower kinetics. The systematic cooling-rate dependence was clearly observed at low temperatures in the inset of Fig. 2B, where the more rapid the cooling rate, the lower the resistivity. Since the equilibrium CCDW state has the highest resistivity, this deviation indicates that the nano-system goes further away from its equilibrium state to metastable states upon rapid cooling.

Taking the kinetics into account, we obtained a more generalized phase diagram as shown in Fig. 2C, displaying the distinct shift of the critical thickness for the occurrence of the NCCDW-CCDW transition as a function of cooling rate. This phase diagram enables us to deduce the critical cooling rate ($R_c$), above which the transition is kinetically avoided upon cooling. Figure 2D is the ground state phase diagram of 1T-TaS$_2$ on the thickness-cooling rate plane. The $R_c$ shows a positive correlation with thickness.

Here, we can compare the kinetics of 1T-TaS$_2$ nano-thick crystals with other materials. For instance, a simple extrapolation indicates that $R_c$ of a 10-nm-thick crystal should be around $10^{-2}$ K/min or less, which is as slow as that of an archetypal good glass former silicon dioxide SiO$_2$ (*24*). It is surprising that such a considerable slowdown of kinetics can be realized just by thinning. Discrete systematic variations in the kinetics were recently reported in organic conductors by substituting ions (*25*). The 1T-TaS$_2$ nano-thick crystal has higher controllability because we can continuously tune the ordering kinetics of the phase transition simply by changing thickness.

**Non-volatile phase switching in 1T-TaS$_2$ nano-thick crystals**

Such ordering kinetics underlies the non-volatile switching in phase-change materials between a quenched glassy state and an equilibrium crystalline state by annealing and quenching (*3,4*), giving us an opportunity to control the first-order phase transition electrically. We applied in-plane voltage on a 24-nm-thick crystal with length 4.8 μm in the super-cooled NCCDW state at 95 K, and realized a switching operation. Blue curves in Fig. 3A and 3B are the voltage dependence of current and sheet resistance ($R_s$), where the current started to decrease gradually above 3 V, followed by the stabilization of a low-current state. Consequently we observed appearance of an insulating behavior of the CCDW phase upon cooling, which is shown as a blue curve in Fig. 3C. This voltage-driven transformation can be a "RESET" process, where the local temperature is increased by the Joule heating due to the current flow, and the crystallization, or growth of CCDW domains, proceeds. Here the local temperature might not exceed the CCDW-to-NCCDW transition temperature.

**Non-volatile switching to a metastable state in 1T-TaS$_2$ nano-thick crystals**

The inverse switching, so called "SET" process, was also achieved. We applied voltage at 165 K on the crystal in the CCDW phase. As shown in the red curves in Fig. 3A and 3B, current suddenly jumped at 3 V and a high-current state was stabilized. Based on the operation mechanism of phase-change memory, this jump of current and $R_s$ may also result from Joule heating, which increases the local temperature above the CCDW-to-NCCDW transition temperature in contrast to the "RESET" process. However, the consequent low-resistivity state was no longer identical to the initial NCCDW state. As shown in Fig. 3C, $R_s$ at 165 K after the application of voltage (red curve) was lower than that of the initial NCCDW state (green curve). Moreover, the $R_s$-$T$ curve never followed the initial curve upon cooling and a metallic conduction was realized below 100 K. The metallic ground state was so robust that it persisted for at least twelve hours at 2 K. Since the conventional phase-change memory is a switching device between two states consisting of low-temperature crystalline and high-temperature glassy phases, our observation of the voltage-induced non-volatile third state implies that the 1T-TaS$_2$ nano-thick crystal is not a simple phase-change material. We may, therefore, need to consider another mechanism besides Joule heating to understand the non-volatile switching behavior.

On the warming up scan, the $R_s$-$T$ curve finally coincided with the initial curve at a certain temperature $T^* = 230$ K, and the system was completely recovered to the initial ICCDW phase. The deviation between the green and red $R_s$-$T$ curves in Fig. 3C is qualitatively similar to what were observed upon rapid cooling shown in Fig. 2, implying that the voltage-induced metallic state (red curves in Fig. 3C) can be also a metastable NCCDW state. Metallic conduction is realized in pressurized or doped 1T-TaS$_2$ systems (*20, 23*), but the existence of such a "metastable" metallic state has never been reported in any 1T-TaS$_2$ system. Here, it was revealed that the application of in-plane voltage is an effective way to induce and stabilize a thermally inaccessible novel non-equilibrium state in 1T-TaS$_2$ nano-thick crystals.

**Memristive phase switching in 1T-TaS$_2$ nano-thick crystals**

The application of voltage under various conditions revealed the memristive nature of 1T-TaS$_2$ nano-thick crystals. Figure 4A is the voltage dependence of $R_s$ in another 24-nm-thick crystal measured at 90 K, where the crystal was initially in a super-cooled NCCDW state. The $R_s$ slightly increased and a relatively insulating state was stabilized after a forward and backward voltage scan. Repeated voltage scans resulted in the gradual increase of $R_s$, followed by the saturation to the equilibrium CCDW state after five cycles. We emphasize that the $R_s$ values near $V = 0$ V and corresponding $R_s$-$V$ curve did not change as long as the applied voltage was limited less than 2 V, indicating that the final state after each cycle is non-volatile and is tuned by the number of cycles. As summarized in Fig. 4B, the $R_s$ depends on the history of the voltage application, demonstrating that the system acts as a "resistor with memory" or memristor (*1,2,5-9,16*). Since most of the memristors utilize the tunnel junction configuration of oxide thin films and heterostructures (*5,6,8,9*), the memristive behavior in a lateral structure (*7,16*) based on a correlated 2D crystal is quite unique. By analogy to the semiconducting NCCDW state in the pressurized 1T-TaS$_2$ (*26*), the increase of $R_s$ may microscopically correspond to the growth of CCDW domains, which is schematically depicted in the insets of Fig. 4B.

The multiple resistive behavior, *i.e.* the memristive characteristic, was also found by applying voltage at higher temperatures. We first cooled down the 24-nm-thick crystal to 90 K and switched from the super-cooled state to the CCDW state by the application of voltage as shown in Fig. 4A. Sequentially we warmed up to 150 K, applied voltage on the crystal in the CCDW state, and obtained conducting $R_s$-$T$ relations (blue curve in Fig. 4C). As in the same way, we observed metallic behavior by applying voltage on the crystal in the CCDW state at 165 K (red curve in Fig. 4C). Moreover, we found that such a conducting state could be also induced by the application of voltage not on the CCDW but on the NCCDW state at 165 K (orange curve in Fig. 4C). In all the three conditions, we could observe the recovery to the normal NCCDW state (green curves) upon warming above $T^*$, indicating all the low-temperature states are metastable NCCDW states. Figure 4D shows the relation between $T^*$ and the $R_s$ at 90 K deduced from Fig. 4C, showing that the $T^*$ increases as the $R_s$ at 90 K is lowered. Since the thermodynamically stable CCDW state shows high resistivity, higher $T^*$ is another sign that the system is further away from its thermal equilibrium state.

## Discussion

Figure 4E is a schematic energy diagram with multiple CDW states that can be realized in a 1T-TaS$_2$ nano-thick crystal, where the CCDW domains shrink and the resistance decreases as the system is set further from its equilibrium CCDW state. This diagram presents the potential of 1T-TaS$_2$ memristor, where we can possibly tune the area of domains and switch to the various intermediate resistance states by applying in-plane voltages. Recent theoretical and experimental research on light-induced metastable states in 1T-TaS$_2$ (*21,27*) may also give some insight into the microscopic nature of the voltage-induced metastable CDW states.

It should be stressed that the metallic NCCDW states (shown in Fig. 4C) have never been reached by rapid cooling but are accessible only by the application of voltage. Also, the voltage application has a peculiar trend; the application of voltage at low temperatures (indicated by rainbow-colored arrows in Fig. 4E) resulting in the switching toward the equilibrium CCDW state, whereas applying electric field at higher temperatures (represented by grey arrows in Fig. 4E) always lead the system into metallic NCCDW states. Such highly controllable but nontrivial responses are not only raising fundamental questions on the phase-change memories and memristors in nanoscale, but also offering potential applications of 2D crystals.

Here, we discuss the mechanism of memristive non-volatile memory operation, especially of the "SET" operation in 1T-TaS$_2$ nano-thick crystals. First, we address the depinning of CDW, which has been intensively investigated in quasi-one-dimensional CDW systems: The application of electric field above the threshold voltage ($V_{th}$) drives the CDW out of the pinned configuration, and consequently realizes switching to a non-linear conduction state (*28,29*). Such a non-linear conduction is reported also in 1T-TaS$_2$ bulk single crystal (*30*). However, the 1T-TaS$_2$ bulk crystal and the nano-thick crystal in the present work are quite different in $V_{th}$. In the nano-thick crystal, we observed that $V_{th}$ was around 3 V (See Fig. 3A), corresponding to 6 kV/cm. On the other hand, $V_{th}$ for bulk single crystals is below 10 V/cm (*30*), which is three orders of magnitude lower than that observed in this work. More important difference between the nano-thick and bulk crystals is in the volatility of the switched state: The final high-current state was kept after the release of the voltage in the nano-thick crystals, whereas in bulk crystals the non-linear conduction vanishes after reducing the voltage. Based on these differences in $V_{th}$ and non-volatility, we conclude that the observed switching behavior in 1T-TaS$_2$ nano-thick crystals cannot be attributed to the depinning of CDW.

Recently, a carrier driven collapse of the Mott gap upon applying voltage is proposed for the mechanism to realize phase switching in 1T-TaS$_2$ (*31*). However, the voltage-induced high-current state is reported to be a volatile ICCDW phase. Existing scenarios hence do not explain the non-volatile switching behavior observed in the present work, indicating that an exotic switching mechanism is working in the 1T-TaS$_2$ nano-thick crystals.

The present results highlight the robustness of the novel metastable states in 1T-TaS$_2$ nano-thick crystals induced by in-plane electric field: While 1T-TaS$_2$ bulk single crystals show volatile memory operations (*30*), we observed non-volatile switching behavior in 1T-TaS$_2$ nano-thick crystals. The sharp contrast between the bulk and nano-thick 1T-TaS$_2$ systems may be in the speed of the ordering kinetics. As clarified in the cooling-rate versus thickness phase diagram (Fig. 2D), the 1T-TaS$_2$ nano-thick crystals with reduced thickness exhibit extremely slow kinetics. We applied in-plane voltage on nano-thick crystals thin enough to realize super-cooled NCCDW states at normal cooling rate 1 K/min. Therefore, non-equilibrium states are easily stabilized in the nano-system, and the voltage-induced metastable metallic states can be so robust that non-volatile memory operation is observed. The current results indicate that the thinning-induced slow kinetics possibly plays a crucial role in stabilizing the metastable states, *i.e.* the non-volatile memory operation.

In conclusion, we demonstrated that the kinetics of the first-order phase transitions in 1T-TaS$_2$ nano-thick crystals can be systematically controlled by changing thickness, and discovered novel metallic metastable states by applying in-plane voltage. The robust various metallic metastable states, which have never been accessed or stabilized in bulk single crystals, are possibly the consequence of the extremely slow kinetics realized by reducing thickness of the crystal to nanometers. Furthermore, we found a memristive function to switch between multi-step non-volatile metastable states. Our findings present a new paradigm that exotic phenomena are hidden not only in 2D crystals with unconventional electronic band structures but also in correlated 2D crystals.

## Methods

**Device preparations.** The 1T-TaS$_2$ mother single crystal was grown by the conventional chemical vapor transport method with 2% excess sulfur.

Nano-thick crystals were isolated from the bulk single crystal by mechanical exfoliation with Scotch tape, and transferred onto doped silicon wafer covered with a layer of thermally grown silicon dioxide. The typical size of cleaved nano-thick crystals was $10 \times 10$ μm$^2$. Metal contacts were made by an electron beam lithography process, followed by the sequential deposition of titanium (5 nm) and gold (100 nm). The thickness of the nano-thick crystals was determined by atomic force microscopy.

**Measurements.** All the transport measurements were performed in Physical Property Measurement System (PPMS, Quantum Design, Inc.) under He-purged conditions. In the cooling-rate effect measurement, when we cooled down and warmed up with the sweeping rate of 1 K/min or smaller, we maintained the rate in the whole temperature scan. When we measured with the sweeping rate larger than 1 K/min, we cooled down to $T = 50$ K with the rapid rate, waited at $T = 50$ K until the sample temperature was stabilized to $T = 50$ K, cooled down and warmed up between $T = 50$ K and $T = 2$ K at 1 K/min, and warmed up with the rapid rate.

**Acknowledgments:** We are grateful to F. Kagawa for fruitful discussions, and thankful to J. T. Ye and T. Iizuka for experimental supports. This work was supported by Grants-in-Aid for Scientific Research (Grant No. 25000003 and 25708040) by the Japan Society for the Promotion of Science (JSPS). Y.I. was supported by the Strategic International Collaborative Research Program (SICORP-LEMSUPER) of the Japan Science and Technology Agency. M.N. was partly supported by Grant for Basic Science Research Projects from The Sumitomo Foundation. M.Y., R.S. and Y.J.Z. was supported by JSPS through a research fellowship for young scientists. **Author contributions:** M.Y. and Y.J.Z. fabricated the devices, and M.Y. performed measurements and analyzed the data. R.S. grew the single crystal. M.Y., M.N. and Y.I. planned and supervised the study. M.Y. and Y.I. wrote the manuscript. **Competing interests:** The authors declare no competing financial interests.

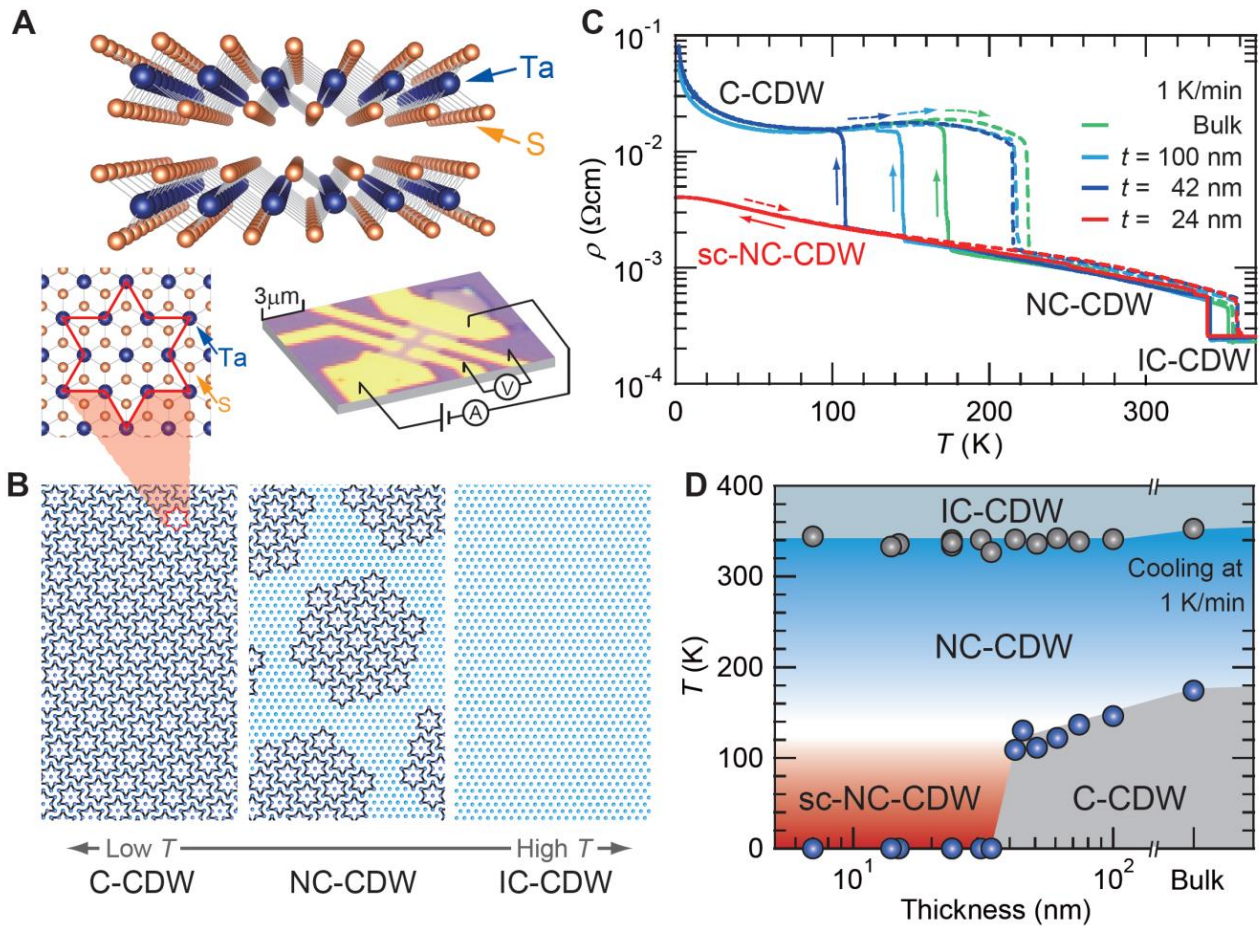

**Fig. 1. 1T-TaS$_2$ single crystals with reduced thickness.** (**A**) The crystal structure of the layered 1T-TaS$_2$, where the planes of tantalum (Ta) atoms are surrounded by sulfur (S) atoms in an octahedral arrangement. The top view of the crystal structure shows a David-star cluster, where twelve Ta atoms within the layer move towards a thirteenth central Ta atom. Also shown is the optical microscope image of a typical nano-thick crystal device. (**B**) The schematic pictures of Ta atom network in the CCDW (left), hexagonal NCCDW (middle) and ICCDW (right) phases. The dark blue circles represent the Ta atoms displaced from their undistorted lattice coordinates, forming the David-star clusters. (**C**) The temperature ($T$) dependence of the resistivity ($\rho$) for bulk and nano-thick crystals of 1T-TaS$_2$. The solid and broken lines represent the $\rho$ in the cooling and warming cycle, respectively. The notation sc-NC-CDW represents super-cooled NCCDW. (**D**) The temperature-thickness phase diagram of 1T-TaS$_2$ nano-thick crystals upon cooling at 1 K/min.

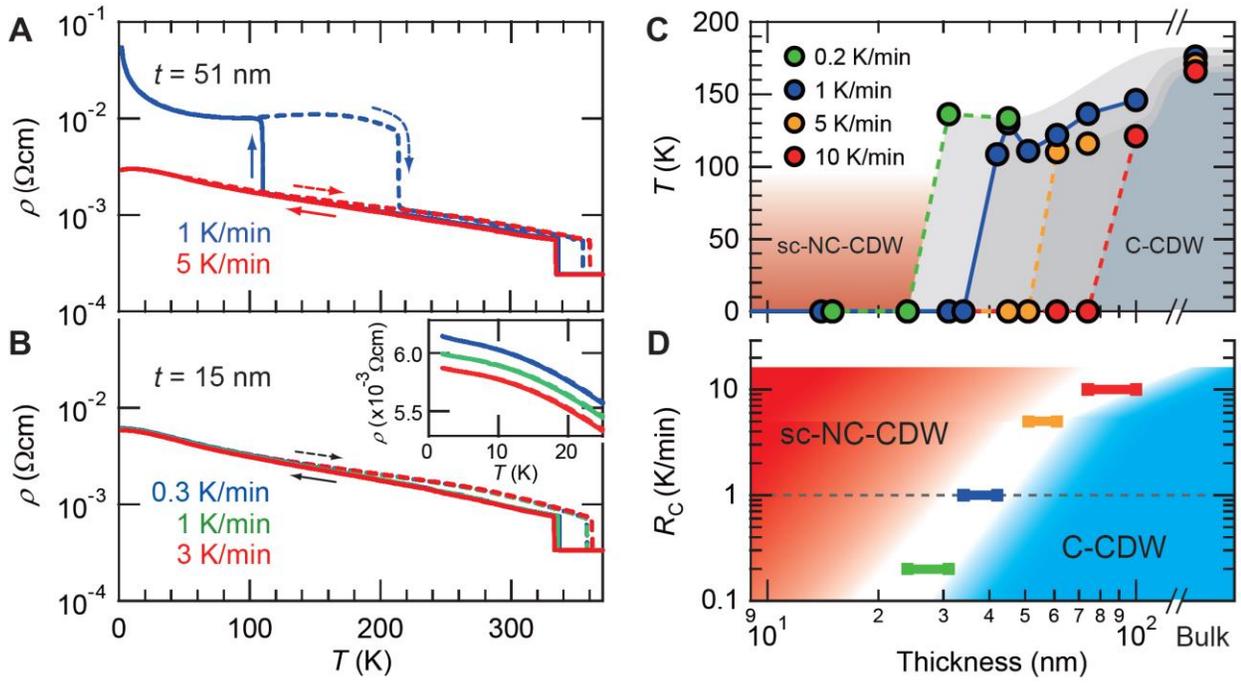

**Fig. 2. Cooling-rate dependent behavior of 1T-TaS$_2$ nano-thick crystals.** (**A** and **B**) The temperature (*T*) dependence of the resistivity (ρ) for different temperature-sweeping rates in crystals with the thicknesses of 51 nm (**A**) and 15 nm (**B**). The inset in **B** is the magnified view of the ρ -*T* curves at low temperatures. (**C**) The temperature-thickness phase diagram of 1T-TaS$_2$ nano-thick crystals for different temperature-sweeping rates. (**D**) The critical cooling rate ($R_c$) versus thickness phase diagram.

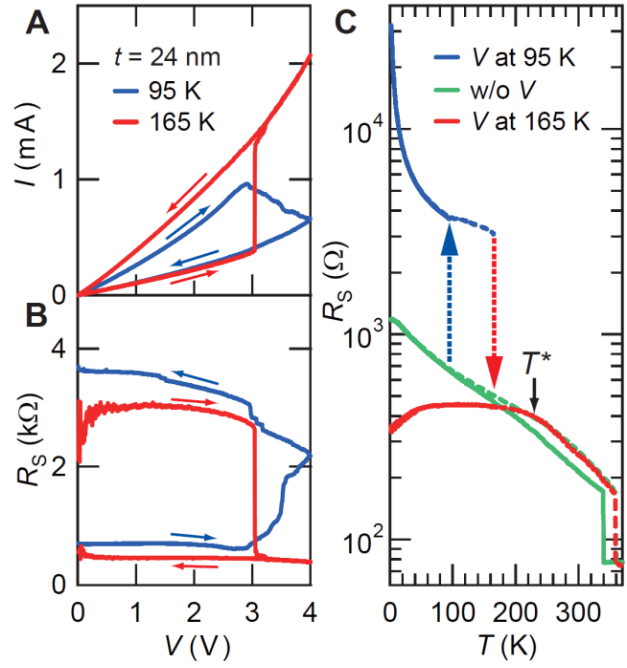

**Fig. 3. Voltage-driven phase switching in a 1T-TaS2 nano-thick crystal.** (**A** and **B**) Current–voltage (*I-V*) characteristics (**A**) and simultaneously measured $R_s$ versus voltage curves (**B**) of a 24-nm-thick crystal measured at 95 K (blue lines) and 165 K (red lines). The sweeping rate was 10 mV/sec at 95 K and 50 mV/sec at 165 K, respectively. (**C**) The temperature dependence of $R_s$ before and after the applications of in-plane voltage. Green curve was taken before the application of voltage. After applying voltage at 95 K, the crystal was cooled down to 2 K and warmed up to 165 K, which data is shown as blue curve. Subsequently we applied voltage on the crystal at 165 K, and cooled and heated the crystal, which is shown as red curves. *T\** is the temperature above which the crystal recovers to its normal NCCDW phase.

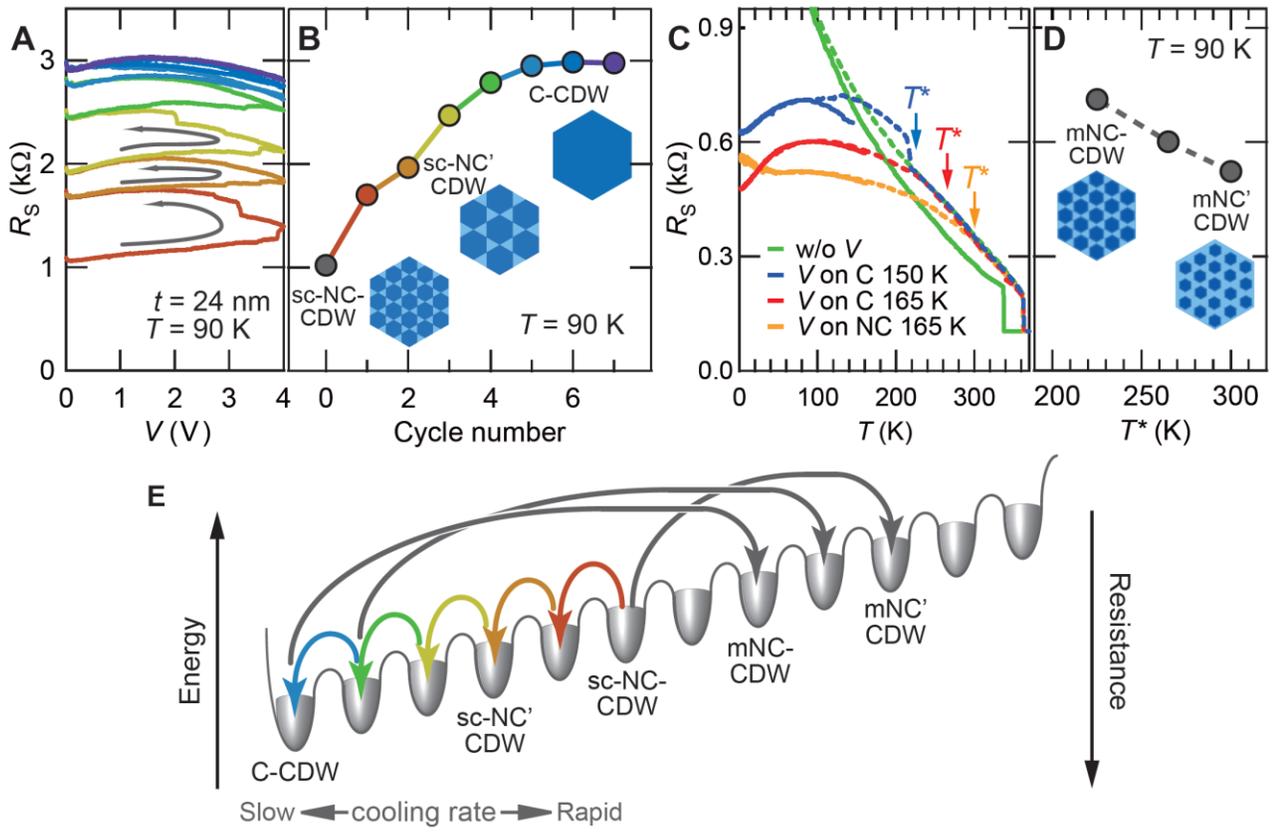

**Fig. 4. Memristive characteristics in a 1T-TaS$_2$ nano-thick crystal.** (**A**) $R_s$ versus voltage (*V*) curves of another 24-nm-thick crystal measured at 90 K with the scanning rate of 10 mV/sec. (**B**) Variation of $R_s$ near $V = 0$ V at 90 K as a function of the number of cycles. Insets are the possible microscopic structures of each state. The dark blue hexagons refer to CCDW domains composed of David-star clusters and the light blue areas represent the discommnesurations. (**C**) The $R_s$-$T$ relations before and after the applications of voltage. Green curve was taken before applying voltage. First the crystal was switched from the super-cooled NCCDW state to the CCDW state by applying voltage at 90 K (shown in **A**) and heated to 150 K. Blue curve is the result of cooling and warming measurements after the application of voltage on the crystal in the CCDW phase at 150 K. Red curve is the results of applying voltage on the crystal in the CCDW phase at 165 K. The orange curve was measured after the voltage application on the crystal in the NCCDW phase at 165 K. (**D**) Variation of $R_s$ near $V = 0$ V at 90 K as a function of $T^*$, above which the system recovers to its normal NCCDW state. Insets are the possible patterns of each metastable state. The notation mNC-CDW represents metallic NCCDW. (**E**) Schematic energy diagram of a 1T-TaS$_2$ nano-thick memristive system, showing a multi-minimum potential. Rainbow- and grey-colored arrows indicate the application of voltage at low and high temperatures, respectively.